# Unravelling Dzyaloshinskii–Moriya interaction and chiral nature of Graphene/Cobalt interface

*Fernando Ajejas [1,2], Adrian Gudín [1], Ruben Guerrero [1], Miguel Angel Niño [1], Stefania Pizzini [3], Jan Vogel [3], Manuel Valvidares [4], Pierluigi Gargiani [4], Mariona Cabero [5], Maria Varela [5], Julio Camarero [1,2], Rodolfo Miranda [1,2], and Paolo Perna [1,*]*

[1] IMDEA Nanociencia, Campus de Cantoblanco, 28049 Madrid, Spain.

[2] Departamento de Física de la Materia Condensada, Instituto "Nicolás Cabrera" & Condensed Matter Physics Center (IFIMAC), Universidad Autónoma de Madrid, Campus de Cantoblanco, 28049 Madrid, Spain.

[3] Université Grenoble Alpes, CNRS, Institut Néel, 38000 Grenoble, France.

[4] ALBA SYNCHROTRON LIGHT SOURCE, Cerdanyola del Vallès, 08290 Barcelona, Spain.

[5] Departamento de Física de Materiales, Instituto de Magnetismo Aplicado & Instituto Pluridisciplinar, Universidad Complutense de Madrid, Ciudad Universitaria 28040, Madrid, Spain.

* Corresponding author: paolo.perna@imdea.org

Date: 20$^{th}$ March 2018

**A major challenge for future spintronics is to develop suitable spin transport channels with long spin lifetime and propagation length. Graphene can meet these requirements, even at room temperature. On the other side, taking advantage of the fast motion of chiral textures, i.e., Néel-type domain walls and magnetic skyrmions, can satisfy the demands for high-density data storage, low power consumption and high processing speed.**
**We have engineered epitaxial structures where an epitaxial ferromagnetic Co layer is sandwiched between an epitaxial Pt(111) buffer grown in turn onto MgO(111) substrates and a graphene layer. We provide evidence of a graphene-induced enhancement of the perpendicular magnetic anisotropy up to 4 nm thick Co films, and of the existence of chiral left-handed Néel-type domain walls stabilized by the effective Dzyaloshinskii–Moriya interaction (DMI) in the stack. The experiments show evidence of a sizeable DMI at the gr/Co interface, which is described in terms of a conduction electron mediated Rashba-DMI mechanism and points opposite to the Spin Orbit Coupling-induced DMI at the Co/Pt interface. In addition, the presence of graphene results in: *i)* a surfactant action for the Co growth, producing an intercalated, flat, highly perfect *fcc* film, pseudomorphic with Pt and *ii)* an efficient protection from oxidation. The magnetic chiral texture is stable at room temperature and grown on insulating substrate. Our findings open new routes to control chiral spin structures using interfacial engineering in graphene-based systems for future spin-orbitronics devices fully integrated on oxide substrates.**

The demand for high density, low power and fast spin logic devices requires the combination of materials that can provide both suitable spin transport channels with long spin lifetime and long-distance spin propagation, and topologically stable spin textures that can act as fast information carriers (1). These prerequisites for the development of spintronic technology in next years can be fulfilled by exploiting the spin and momentum degrees of freedom of electrons (spin-orbitronics) (2) in specific superlattice structures.

On one side, graphene (gr), a single atomic layer of graphite, is a promising spin channel material owing to the achievement of room temperature (RT) spin transport with long spin diffusion lengths of several micrometers (3). However, the development of gr-spintronic devices requires that other active properties are incorporated to graphene. The generation of long range magnetic order and spin filtering in graphene have been recently achieved by molecular functionalization, as well as the introduction of giant spin-orbit coupling (SOC) in the electronic bands of graphene by intercalation (4)(5). Graphene may also affect the tunnel magnetoresistance (6), the spin injection efficiency (7)(8), the Rashba effect (9)(10), or the quantum spin-Hall effect (11) and the perpendicular magnetic anisotropy (PMA) (12)(13). This, together with its passivating properties, inertness and impermeability favors the integration of graphene in spin-orbitronic devices.

On the other side, high processing speed and high dense data storage with reduced energy consumption can be achieved by exploiting chiral Néel-type magnetic domain walls (DWs) (14)(15)(16) and magnetic skyrmions (17)(18) as carriers of digital information. These magnetic structures are obtained in non-centrosymmetric multilayer stacks with PMA and in presence of the antisymmetric Dzaloshinskii-Moriya exchange interaction (DMI) that favors a chiral arrangement of spins within the DWs (19). Such DMI stabilizes also magnetic skyrmions, i.e. the topologically protected, nanometer-sized, whirling magnetic objects which have been observed in the presence of strong magnetic fields and at low temperature (17)(20) and recently detected at RT under low or vanishing magnetic fields (21)(22)(23)(24). By balancing the uniaxial PMA and the DMI by acting on the nature and thickness of the materials composing the stacks, one can tune the stability and the dimension of the chiral Néel DWs (21)(25). In particular, the DMI strength can be tuned by using two-active interfaces (26)(27)(28).

The integration of graphene as efficient spin transport channel in the chiral DWs technology depends on our ability to fabricate gr-

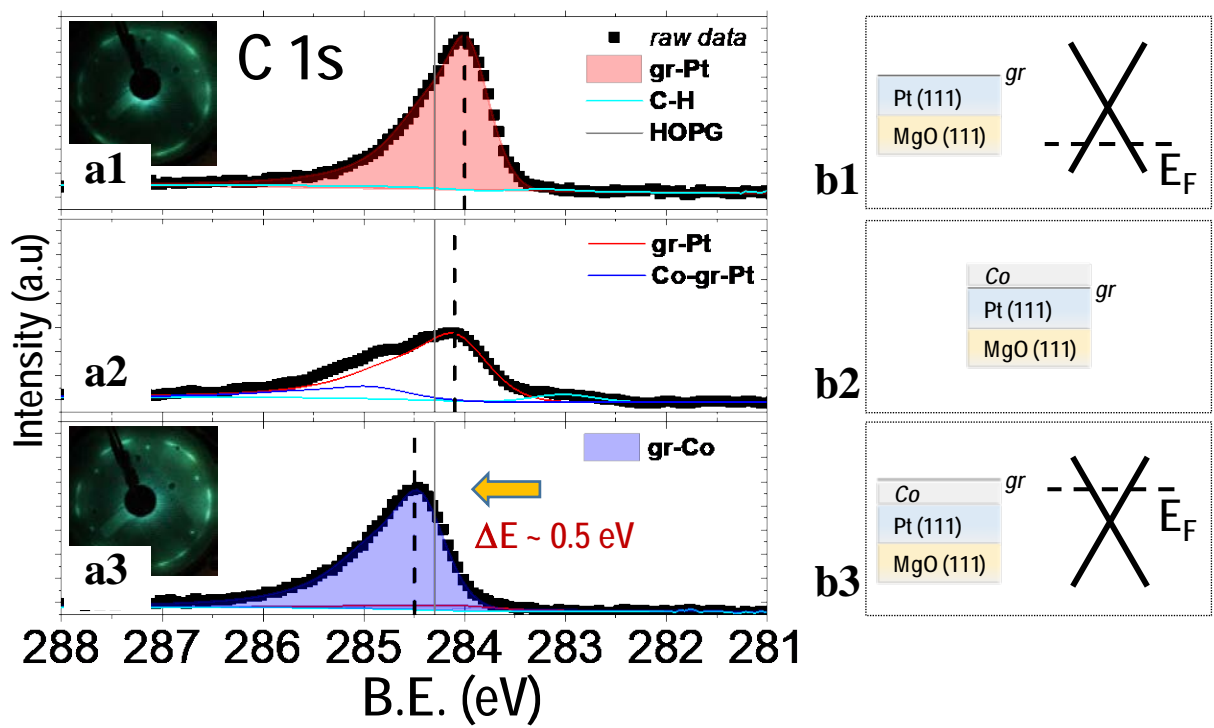

**Figure 1. (a)** XPS-LEED analysis at each stage of the growth process indicated in the right-side sketches together with **(b)** the schematic position of the Fermi energy, $E_F$, with respect to the Dirac cone for *p*-doped (i.e., gr/Pt(111)) and n-doped (i.e., gr/Co(111)) graphene. **(a1)**: the C 1s peak appears at binding energy (BE) = 284.0 eV (red component in the fit) in gr/Pt(111). The LEED image presents the pattern characteristic of multi-domain gr (domains oriented ±15°). **(a2)**: after the Co evaporation (i.e., Co/gr/Pt(111)), a new component of C *1s* ascribed to Co/gr/Pt is found at 285.0 eV (blue component). The component due to gr/Pt shifts of +0.1 eV from the original peak. The Co evaporated on gr forms islands and clusters so that no LEED is obtained. **(a3)**: once the Co is completed intercalated (i.e., gr/Co/Pt(111)), the C *1s* peak shifts towards higher energy by +0.4 eV and is found at 284.5 eV. After Co intercalation, the clean condition of graphene and its corresponding LEED pattern is recovered. The grey line indicates the C *1s* peak position relative to graphite (HOPG).

F. Ajejas *et al.* Unravelling Dzyaloshinskii–Moriya interaction and chiral nature of Graphene/Cobalt interface (2018)  2

based PMA systems with tailored interfacial SOC. So far, the studies on gr-based magnetic systems are rather scarce, and, typically, make use of metallic single crystals as substrates which jeopardize the exploration of their transport properties (since the current is drained by the substrate). Here we report on the realization and characterization of epitaxial asymmetric gr/Co/Pt(111) structures grown on (111)-oriented oxide substrates, displaying at RT enhanced PMA and chiral left-handed Néel-type DWs stabilized by DMI. The PMA, extended up to 4 nm Co films, and DMI originate from SOC at the Pt/Co and the Co/gr interfaces. The exceptional properties of graphene consent to get homogenous, flat and protected magnetic layer, and the use of oxide substrates allows the stacks to be integrated in future spin-orbitronics devices.

In order to realize high quality epitaxial gr-based PMA heterostructures with atomically flat interfaces and a homogeneous ferromagnetic (FM) layer, we have followed the procedure detailed in Methods. It consisted in preparing first 30 nm thick epitaxial Pt(111) buffers grown by dc sputtering on pre-treated commercially available MgO(111) substrates, on which an epitaxial layer of graphene was grown by ethylene chemical vapor deposition (CVD) at 1025 K. The cobalt films, with thickness $t_{Co}$ ranging from 1 to 6 nm, were hence evaporated on top of graphene by molecular beam epitaxy (MBE) at room temperature and low deposition rate. The temperature was gradually raised to less than 550 K in order to activate the Co intercalation process while avoiding unwanted metal intermixing.

The entire process was performed in ultra-high-vacuum (UHV) conditions and monitored by *in-situ* surface analysis, in order to obtain homogeneous epitaxial cobalt layers, sandwiched between a Pt(111)-buffer and graphene.

The excellent surface quality is demonstrated by the low energy electron diffraction (LEED) patterns shown in the left-side insets of **Figure 1a**. The Pt(111)-buffer sputtered on the (111)-oriented oxide substrate presents a crystal quality equivalent to a Pt(111) single crystal. The typical LEED rings due to multi-domain gr flakes (domains oriented ±15°) characterize the gr/Pt(111)/MgO(111) stack (28). Once the Co is successfully intercalated, the LEED pattern of gr/Pt(111) was recovered. These evidences indicate that Co is homogeneously intercalated underneath gr over ~µm², and that there, it grows epitaxially, pseudomorphic with Pt, and (111)-oriented. A representative collection of LEED images acquired at each stage of the fabrication is shown in Figure S1 of the Supplementary Information.

In order to get full control on the chemical and electronic interface properties, *in-situ* x-ray photoemission spectroscopy (XPS) spectra at the C *1s* (**Figure 1**), Co *2p* and Pt *4d* edges (Figure S1) have been acquired at each stage of the process, i.e. gr/Pt, Co/gr/Pt and gr/Co/Pt. **Figure 1** shows the XPS spectra of the C *1s* (panels a) accompanied by illustrations indicating in which stage of the growth they were taken (panels b). In the panels b, sketches of Fermi energy ($E_F$) with respect to Dirac cone for p-doped [i.e., gr/Pt(111)] and n-doped [i.e., gr/Co(111)] graphene are shown. After the growth of gr on top of the epitaxial Pt(111) buffer (**Figure 1a1**) the C *1s* line shape can be fitted with one component at Binding Energy (BE) = 284.0 eV ($C_{1s}$ $^{gr\text{-}Pt}$, filled red area), which corresponds to the sp² hybridization of the C-C bonding on the weakly interacting Pt(111) (30). The C *1s* level in gr/Pt(111) is shifted to lower BE by 0.26 eV with respect to the one of graphite (HOPG) (31). The shift in the C core level reflects the position of the Dirac point with respect to the Fermi level ($E_D$=+0.26 eV), in agreement with ARPES data for gr/Pt(111) (32). Thus, graphene on Pt(111) is p-doped.

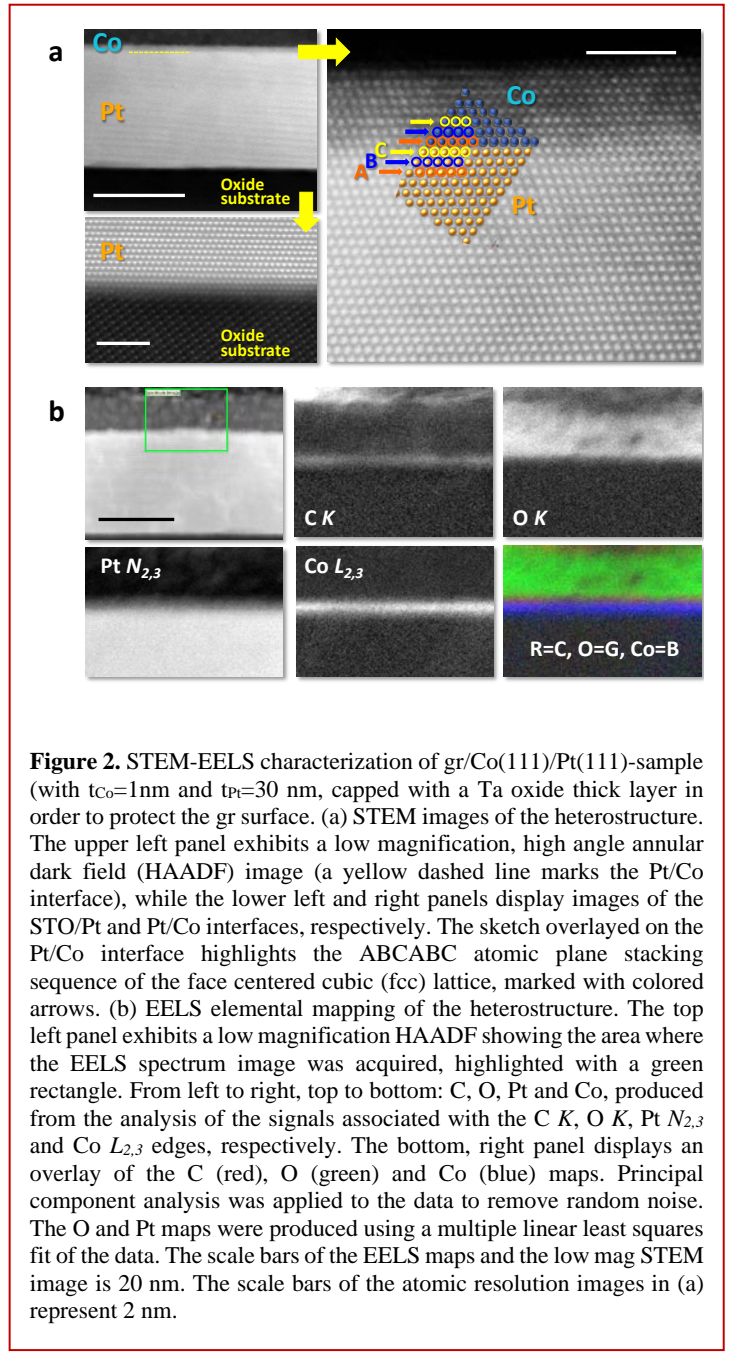

**Figure 2.** STEM-EELS characterization of gr/Co(111)/Pt(111)-sample (with $t_{Co}$=1nm and $t_{Pt}$=30 nm, capped with a Ta oxide thick layer in order to protect the gr surface. (a) STEM images of the heterostructure. The upper left panel exhibits a low magnification, high angle annular dark field (HAADF) image (a yellow dashed line marks the Pt/Co interface), while the lower left and right panels display images of the STO/Pt and Pt/Co interfaces, respectively. The sketch overlayed on the Pt/Co interface highlights the ABCABC atomic plane stacking sequence of the face centered cubic (fcc) lattice, marked with colored arrows. (b) EELS elemental mapping of the heterostructure. The top left panel exhibits a low magnification HAADF showing the area where the EELS spectrum image was acquired, highlighted with a green rectangle. From left to right, top to bottom: C, O, Pt and Co, produced from the analysis of the signals associated with the C *K*, O *K*, Pt $N_{2,3}$ and Co $L_{2,3}$ edges, respectively. The bottom, right panel displays an overlay of the C (red), O (green) and Co (blue) maps. Principal component analysis was applied to the data to remove random noise. The O and Pt maps were produced using a multiple linear least squares fit of the data. The scale bars of the EELS maps and the low mag STEM image is 20 nm. The scale bars of the atomic resolution images in (a) represent 2 nm.

After the evaporation of 1 nm of Co, the C signal decreases, attenuated by Co covering the graphene monolayer and a new component at higher BE appears related to the Co-C bonding (**Figure 1a2**).

By annealing the sample at 550 K during 30 minutes, Co intercalates completely underneath the gr monolayer. The XPS spectrum in the **Figure 1a3** shows a corresponding increase of the intensity of the C 1s peak, which recovers the initial lineshape and is shifted to a higher BE of 284.5 eV, i.e. +0.5 eV higher than the original $C_{1s}^{gr-Pt}$. This indicates that gr/Co is n-doped, with the Dirac point at +0.14 eV in agreement with the respective metal work function ($\phi_{Co}$ = 5.0 eV, $\phi_{Pt}$ 5.93 eV) (31). This picture is also consistent with ARPES experiments for Co intercalated in gr/Ir(111) (33). The intercalation process, if carried out in the temperature window described here, does not produce intermixing of Co and Pt, as verified by XPS (see Figure S2 in Supplementary Information) and EELS (see below).

These experimental findings demonstrate that graphene is p-doped when grown onto Pt(111), and n-doped on Co (side sketches in Figure 1b). The observed shift demonstrates that gr acquires charge from the Co film underneath and accounts therefore for the presence of an electric field gradient at the gr/Co interface.

The superior quality of the stacks has been proven also ex-situ by high-resolution aberration-corrected scanning transmission electron microscopy (STEM) and electron energy loss spectroscopy (EELS) measurements. Low magnification images exhibit layers that are flat and continuous over lateral distances in the micron range (top left image in **Figure 2.a**). The atomic resolution STEM images of the Pt/Co interface demonstrate a clear ABC-type stacking within both Co and Pt layers, characteristic of a well ordered *fcc* structure and reveal the coherent growth and absence of major defects or dislocations at the Pt/substrate, Co/Pt and C/Co interfaces (right and bottom images in Figure 2a). Within the sensitivity of the technique, the Co layer is pseudomorphic, fully strained in plane to match the underlying Pt buffer. The compositional maps in **Figure 2b**, obtained from EELS, show the absence of major Co/Pt intermixing and that the continuous graphene layer acts as an effective capping to prevent Co oxidation when removing the samples from the growth chamber. Noteworthy, these are the first STEM images acquired in gr-based magnetic heterostructures.

This growth methodology allows getting full control on the chemical and electronic properties of the different interfaces composing the stacks fully integrated on oxide substrates. In addition, it ensures superior crystal quality of the homogeneous layers displaying *fcc* structure and atomically flat interfaces.

The magnetic properties of the stacks were investigated at RT by both averaging and imaging techniques. **Figure 3.a** shows selected *perpendicular* (polar Kerr) hysteresis loops acquired in samples with different Co thicknesses ($t_{Co}$) demonstrating that graphene-covered Co(111)/Pt(111) structures still reveal PMA up to Co thickness of 4 nm (20 ML). For the thinnest Co film investigated, i.e., 5 MLs, the *in-plane* (VSM-SQUID) hysteresis loop displays a reversible smooth transition (**Figure 3b**). i.e., hard-axis behavior, with saturation magnetization $M_S \sim 1.3$ MA/m and very large anisotropy field $\mu_0 H_K = 2$T. Up to 15 ML of Co, the perpendicular hysteresis loops show sharp transitions, large coercive field, and fully remanent state (**Figure 3a,** panels a1-a3). As the Co thickness increases, more rounded transitions accompanied by smaller coercive fields and lower remanence magnetization are observed (panels a4-a6). This indicates a progressive decrease of the magnetic anisotropy with increasing Co thickness, which is confirmed by the evolution of the remanence magnetization ($M_{z,rem}$) normalized to the saturation magnetization ($M_S$) (**Figure 3c**). It proves that the magnetization points essentially out-of-plane up to 20 MLs of Co. In the case of polycrystalline Pt/Co(t)/Pt symmetric stacks, the out-of-plane to in-plane anisotropy critical transition occurs much earlier, i.e., around 7 MLs (solid line of Figure 3b). The critical thickness for reorientation transition found in such oxide-integrated epitaxial gr/Co/Pt(111) is the largest one reported to date (12).

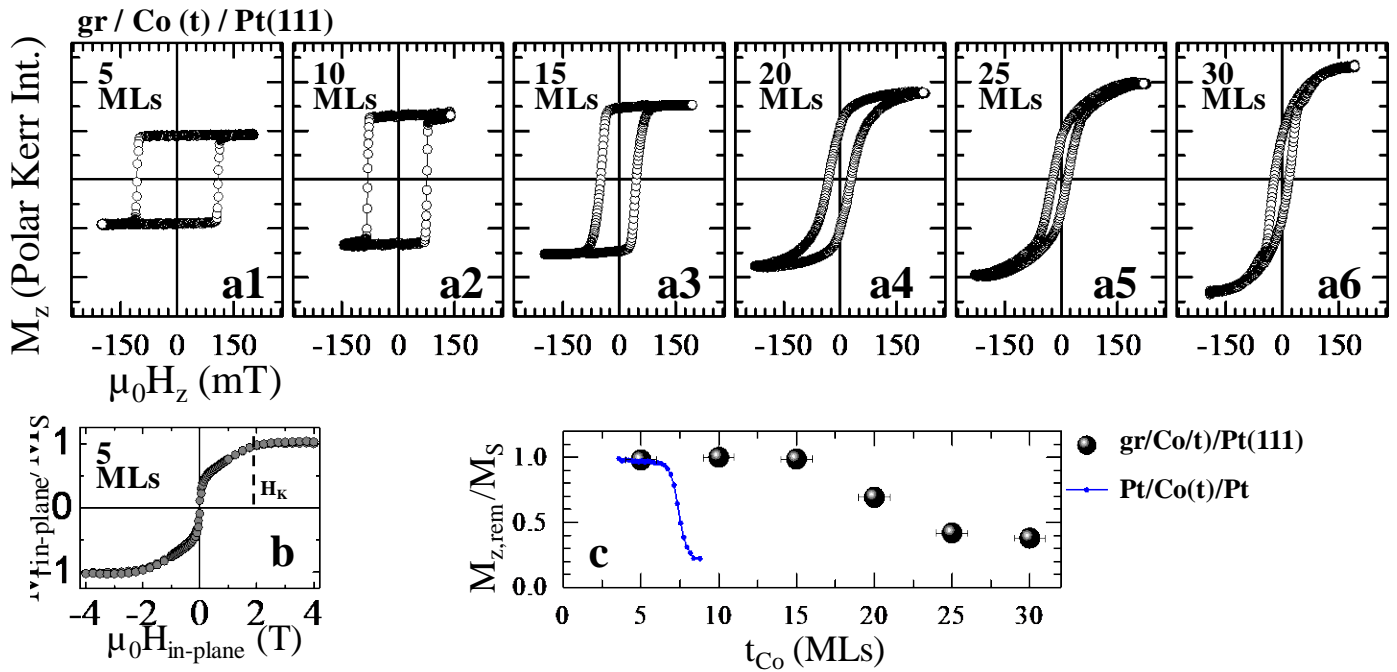

**Figure 3.** **(a)** Perpendicular (polar Kerr) hysteresis loops of gr/Co(t) over Pt(111)-buffers for the indicated Co thickness. **(b)** In-plane (VSM-SQUID) hysteresis loop of the 5MLs thick Co sample, from which the anisotropy field results $\mu_0 H_K$=2T and the saturation magnetization $M_S$=1.3 MA/m. **(c)** Evolution of the perpendicular remanence magnetization normalized to the saturation magnetization ($M_{z,rem}/M_S$) as function of the Co thickness. Black dots refer to the epitaxial gr-based systems, i.e., extracted from the loops shown in (a), while blue-line corresponds to polycrystalline Pt(4nm)/Co(t)/Pt(10nm) grown onto Si-SiO$_2$ substrates.

The microscopic origin of the large PMA has been disclosed by synchrotron-based x-ray absorption (XAS) and magnetic circular dichroism (XMCD) measurements. **Figure 4** compares RT spectroscopic data at the Co $L_{2,3}$ edges obtained in perpendicular (panel a) and grazing (panel b) incidence geometries, as schematically shown in the insets. First, a clear metallic XAS line shapes and XMCD can be observed, proving that the Co atoms are not oxidized after exposure to air because of the effective protection of the graphene ML on top. Second, in perpendicular geometry, the identical dichroic signal found at saturation (symbols) and remanence (shadowed area) confirms the well-defined PMA. Third, the large difference of the dichroic signals measured in perpendicular and grazing conditions indicates large PMA, as discussed below.

Quantitative information can be obtained by using the XMCD sum rules (34), i.e., the projections of the orbital ($m_L$) and spin ($m_S$) magnetic moments along the incident light direction (see Methods and Figure S3). The absolute $m_S$ and $m_L$ values may not be compared with other experiments and theory since their determination depends on the number of holes $n_h$ in he Co *3d* states, which is difficult to estimate, and in addition, may change considerably due to the electric polarization. However, the ratio $m_L/m_S$ and the normalized to hole values of the magnetic moments are significant for comparison since they do not suffer from such problems. In the 5 MLs thick Co sample, the ratio along the anisotropy axis, $m^\perp_L/m^\perp_S = 0.15$, is larger than the one found at grazing incidence, i.e. $m^\parallel_L/m^\parallel_S = 0.11$ which is closer to the *hcp* Co bulk value ($m^{bulk}_L/m^{bulk}_S = 0.10$ (34)). The large $m^\perp_L/m^\perp_S$ is due to the enhancement of the perpendicular orbital momentum. The derived moments at RT are $m^\perp_L/n_h = 0.10$ $\mu_B$/hole and $m^\parallel_L/n_h = 0.06$ $\mu_B$/hole ($\sim m^{bulk}_L/n_h = 0.06$ $\mu_B$/hole (34)). In other words, the orbital moment is strongly confined in the sample normal direction. Remarkably, $m^\perp_L$ is larger than any other values reported for *hcp* Co (35), whereas it is close to the one found for two-dimensional Co islands on Pt(111) by Gambardella *et al.* at low temperature (i.e., $m_L/n_h = 0.12$ $\mu_B$/hole at 10 K (36), and 0.13 $\mu_B$/hole at 5.5 K (37)). The enhanced $m^\perp_L$ is responsible of the large SO anisotropy, $\Delta E_{SO}$, proportional to $\Delta m_L/n_h = (m^\parallel_L - m^\perp_L)/n_h$ (38). In this case $\Delta m_L/n_h = -0.045$ $\mu_B$/hole, which corresponds to -0.11 $\mu_B$/atom assuming $n_h = 2.49$ (34). First principle calculations suggest that the large anisotropy is due to the hybridization of the Co $d_{yz}$ and $d_{z^2}$ orbitals with the $\pi$ states of gr (12)(33). These measurements prove that the PMA of Co(111) films sandwiched between gr and Pt(111) is due to the large anisotropy of the orbital moment.

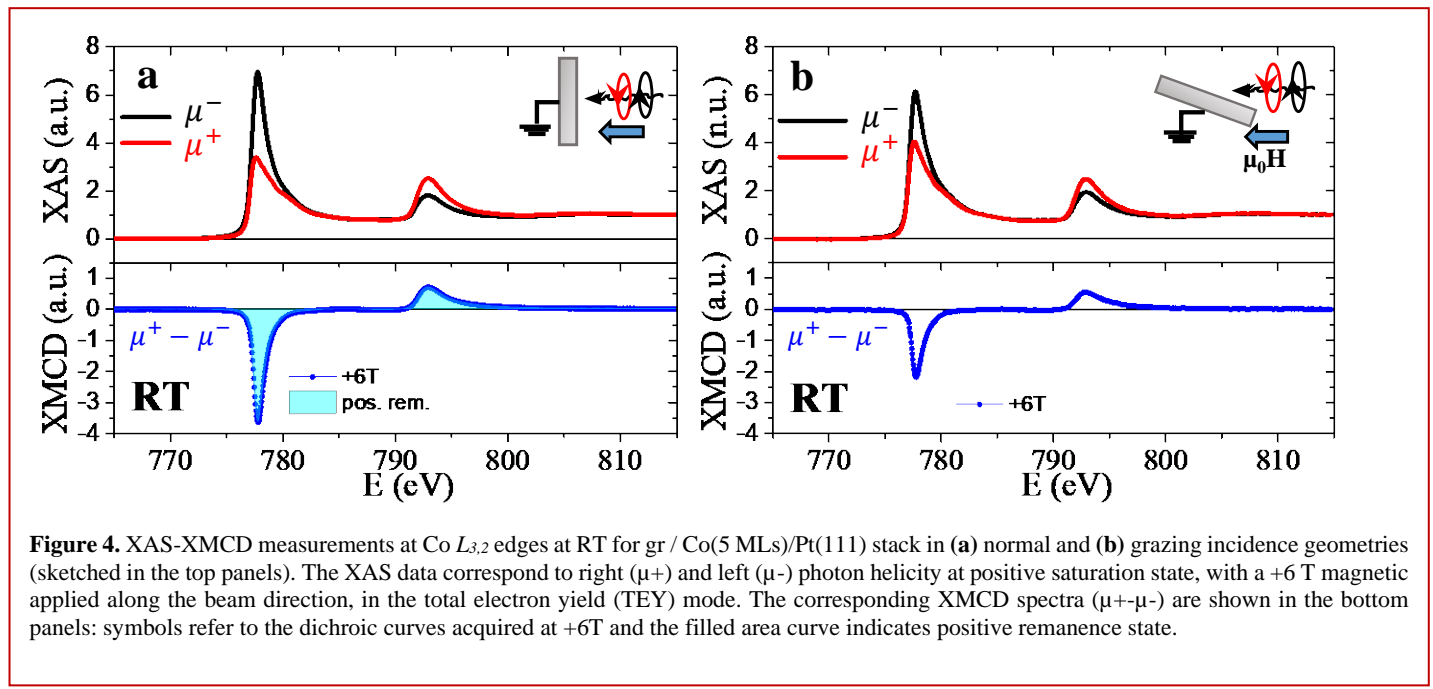

**Figure 4.** XAS-XMCD measurements at Co $L_{3,2}$ edges at RT for gr / Co(5 MLs)/Pt(111) stack in **(a)** normal and **(b)** grazing incidence geometries (sketched in the top panels). The XAS data correspond to right ($\mu+$) and left ($\mu-$) photon helicity at positive saturation state, with a +6 T magnetic applied along the beam direction, in the total electron yield (TEY) mode. The corresponding XMCD spectra ($\mu+$-$\mu-$) are shown in the bottom panels: symbols refer to the dichroic curves acquired at +6T and the filled area curve indicates positive remanence state.

The existence of chiral magnetic textures in the oxide-integrated epitaxial gr/Co/Pt(111) system has been proved by studying the dynamics of the magnetic domain walls (DWs) by Kerr microscopy experiments (28)(39).

**Figure 5a** shows the DW velocity in a gr/Co(5MLs)/Pt(111) stack as function of the intensity of a pulsed out-of-plane magnetic field ($\mu_0 H_z$) (see Methods for details). The velocity increases when entering the flow regime and saturates above 550 mT. This is the behavior expected for a system with non-zero effective DMI, where the DWs acquire chiral Néel internal structure (39)(40).

The effective DM energy density (D) and the sign of the DW chirality can be experimentally determined by exploiting the fact that the $H_z$-driven propagation of chiral Néel walls becomes anisotropic in the direction of an in-plane field applied perpendicular to the domain wall direction ($\mu_0 H_x$) (41). The DMI acts in fact as a local transverse field ($H_{DMI}$) having opposite directions for up/down and down/up DWs. The DW velocity presents hence a minimum when the applied in-plane field compensates the $H_{DMI}$ field. **Figure 5b** shows the velocity vs. $\mu_0 H_x$ curve, which exhibits a minimum for $\mu_0 H_{DMI} = \pm 120$ mT for down/up and up/down DWs respectively. The sign of the $H_{DMI}$ confirms the presence of left-handed DW chirality. On the other hand, D can be deduced from the values of the DMI field, by using the expression $\mu_0 H_{DMI} = \frac{D}{M_S \Delta}$ (19)(39), where $\Delta = \sqrt{A/K_0}$ is the DW width, $M_S$ the saturation magnetization ($M_S \sim 1.3$ MA/m from Figure 3c), A the exchange stiffness (A=24 pJ/m from (42)), $K_0$ the effective anisotropy energy ($K_0 = M_S H_K /2 \sim 1.3$ MJ/m$^3$ from Figure 3c). The estimated DW width is $\Delta=4.1$ nm and the effective DM energy density is $D = 0.6 \pm 0.2 \ mJ/m^2$. By assuming that the value for the DMI at the epitaxial Co/Pt(111) interface is $D^{Co/Pt} = 1.4 \pm 0.2 \ mJ/m^2$ (in Ref. (28)), the decrease of the effective DMI can be attributed to a DMI of opposite sign at the gr/Co interface with a strength of order of $D^{gr/Co} = -0.8 \pm 0.2 \ mJ/m^2$, which corresponds to -0.6 meV/Co-atom.

This scenario is schematically illustrated in **Figure 6**. The DMI at the bottom Co/Pt(111) interface can be explained in terms of the Fert-Levy model (43) involving two magnetic ions (with spins $S_{1,2}$ at position $R_{1,2}$) and a Pt heavy-metal impurity, and produces a canting of the $S_{1,2}$ spins. The DM energy is expressed by $E_{Co/Pt}^{DM} = D_{12} \cdot (S_1 \times S_2)$ where the DM vector $D_{12} \propto (R_1 \times R_2)$ points perpendicular to the triangle formed by the three ions (1) ($D_{12} \| -\hat{y}$). Nevertheless, this model may fail in giving a correct picture of the physics at the gr/Co. Rather, at this interface the observed electric field gradient (Figure 1) that points perpendicular to the sample surface (i.e., $E = E_0 \hat{z}$), gives rise to a Rashba DM term (44)(45). In fact, the electrons confined in the interface plane experience an effective magnetic field that couples with the spin (Rashba-SOC). In the case of FM films in which the magnetization is totally confined along $\hat{z}$, the Rashba-induced DM energy involving the magnetic ions $S_{3,4}$ at position $R_{3,4}$ takes the form $E_{gr/Co}^R = D_{34} \cdot (S_3 \times S_4)$ (44). The $D_{34}$ is proportional to $(\hat{z} \times R_{34})$, and results to be parallel to $\hat{y}$, i.e. opposite to $D_{12}$, in agreement with the experimental results.

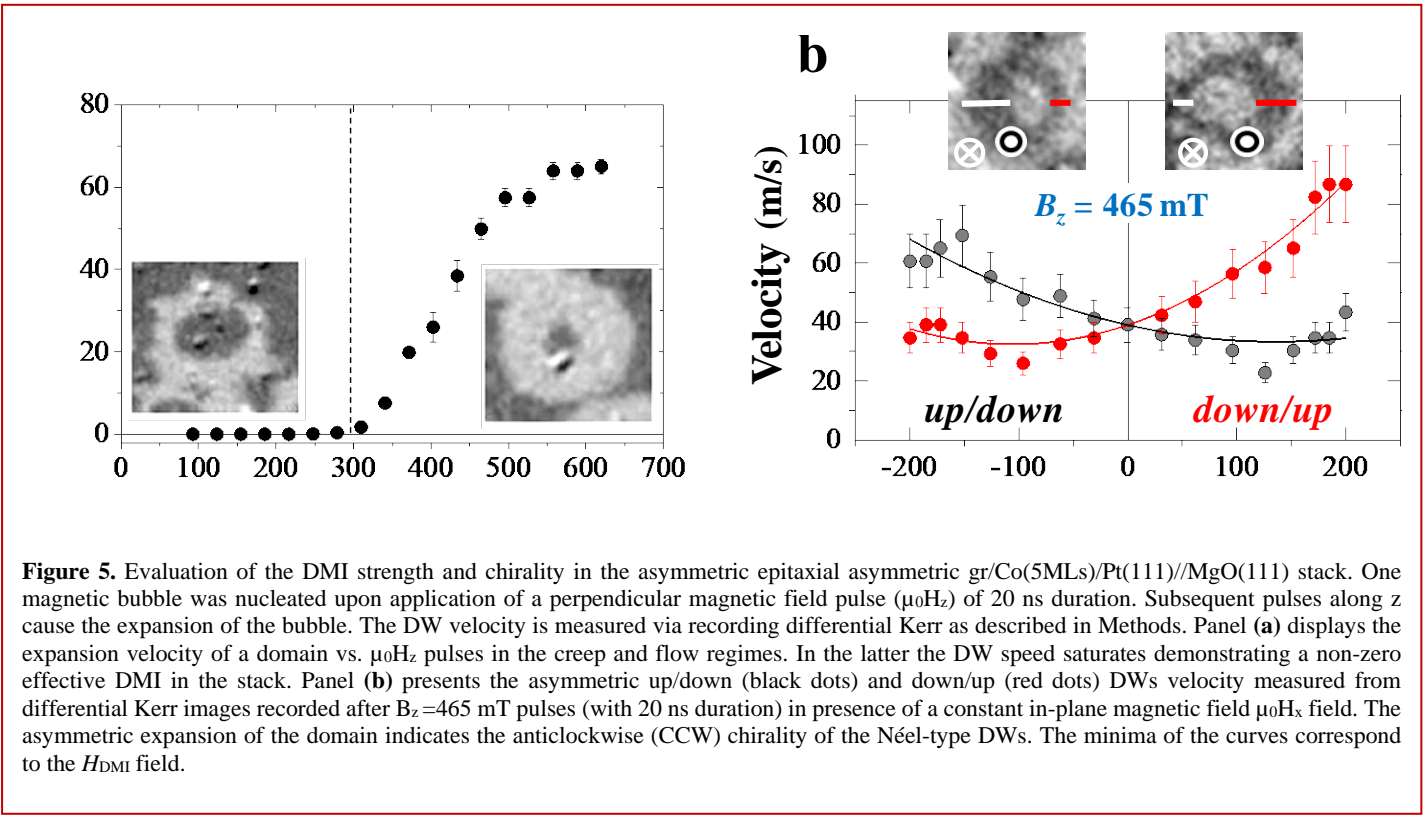

**Figure 5.** Evaluation of the DMI strength and chirality in the asymmetric epitaxial asymmetric gr/Co(5MLs)/Pt(111)//MgO(111) stack. One magnetic bubble was nucleated upon application of a perpendicular magnetic field pulse ($\mu_0 H_z$) of 20 ns duration. Subsequent pulses along z cause the expansion of the bubble. The DW velocity is measured via recording differential Kerr as described in Methods. Panel (**a**) displays the expansion velocity of a domain vs. $\mu_0 H_z$ pulses in the creep and flow regimes. In the latter the DW speed saturates demonstrating a non-zero effective DMI in the stack. Panel (**b**) presents the asymmetric up/down (black dots) and down/up (red dots) DWs velocity measured from differential Kerr images recorded after $B_z$=465 mT pulses (with 20 ns duration) in presence of a constant in-plane magnetic field $\mu_0 H_x$ field. The asymmetric expansion of the domain indicates the anticlockwise (CCW) chirality of the Néel-type DWs. The minima of the curves correspond to the $H_{DMI}$ field.

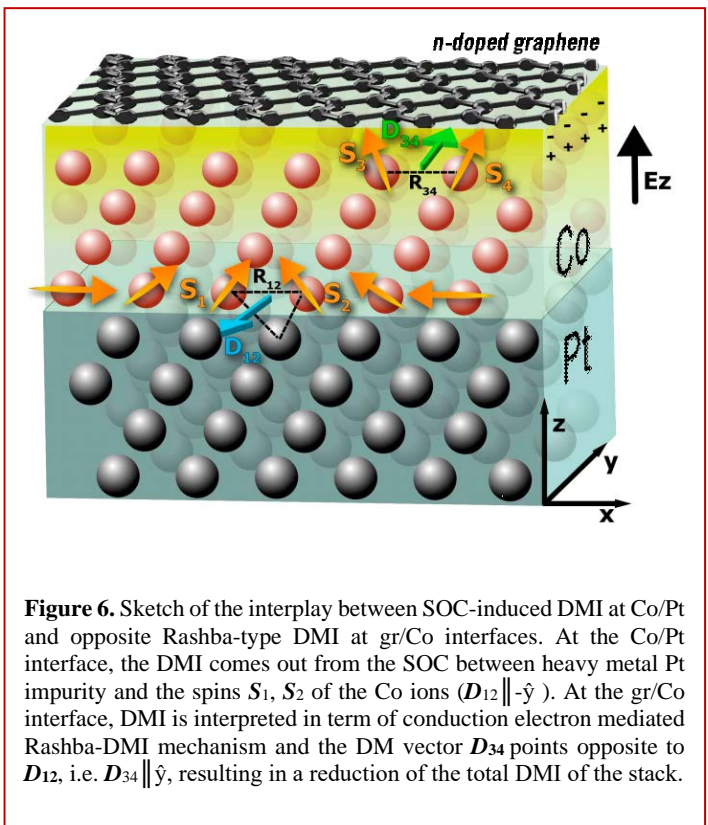

**Figure 6.** Sketch of the interplay between SOC-induced DMI at Co/Pt and opposite Rashba-type DMI at gr/Co interfaces. At the Co/Pt interface, the DMI comes out from the SOC between heavy metal Pt impurity and the spins $S_1$, $S_2$ of the Co ions ($\boldsymbol{D}_{12} \parallel -\hat{y}$). At the gr/Co interface, DMI is interpreted in term of conduction electron mediated Rashba-DMI mechanism and the DM vector $\boldsymbol{D}_{34}$ points opposite to $\boldsymbol{D}_{12}$, i.e. $\boldsymbol{D}_{34} \parallel \hat{y}$, resulting in a reduction of the total DMI of the stack.

## Conclusions

We have demonstrated the ability to engineer epitaxial structures where the ferromagnetic Co layer is sandwiched between an epitaxial Pt(111) buffer grown onto oxide substrates and a graphene layer. We provide evidence of an enhanced PMA, up to 4 nm thick Co films, and of chiral left-handed Néel-type DWs stabilized by the effective DMI in the stack. In the case of 1 nm thick Co layer underneath gr, we found large orbital momentum $m^{\perp}_L = 0.25$ $\mu_B$/atom, PMA of 1.3 MJ/m$^3$ and an effective DMI of around 0.6 mJ/m$^2$. The experiments show evidence of a large DMI at the gr/Co interface, which is described in terms of a conduction electron mediated Rashba-DMI mechanism and points opposite to the SOC-induced DMI at the Co/Pt interface, resulting in a reduction of the total DMI of the stack. In addition, the presence of graphene results in: *i)* a surfactant action for the Co growth, producing an intercalated, flat, highly perfect fcc film, pseudomorphic with Pt; *ii)* an enhancement of the PMA of Co(111)/Pt(111); *iii)* an effective tuning of the DMI that allows the presence of Néel-type chiral magnetic structures; and *iv)* an efficient protection from oxidation by air exposure. The discovery of a strong DMI at the Graphene/Cobalt interface is a crucial step to promote 2D materials spin-orbitronics.

## ACKNOWLEDGMENTS

P.P. acknowledges Prof. A. Fert for fruitful discussions and interesting suggestions. This work has been supported by MINECO (Ministerio de Economía y Competitividad) through FLAGERA Graphene Flagship ('SOgraph' No. PCIN-2015-111) and through Projects FIS2016-78591-C3-1-R (SKYTRON), FIS2015-67287-P, MAT2012-39308, MAT2015-66888-C3-3-R, MAT2014-59315-R, FIS2013-45469-C4-3-R, FIS2016-78591-C3-2-R (AEI/FEDER, UE) and by the Comunidad de Madrid through Project S2013/MIT-2850 NANOFRONTMAG-CM. Financial support from the ERC PoC2015 MAGTOOLS is also acknowledged. IMDEA-Nanociencia acknowledges support from the 'Severo Ochoa' Program for Centres of Excellence in R&D (MINECO, Grant SEV-2016-0686).

## Methods

**Sample preparation.** In order to realize epitaxial gr-based PMA stacks with atomically flat interfaces, we need to consider two main issues regarding the fabrication process: *i)* the synthesis of epitaxial graphene requires high temperature, and this may cause interfacial intermixing; *ii)* the evaporation of metals on graphene typically forms disordered clusters with no defined anisotropy (46)(47). To overcome these problems, the asymmetric epitaxial gr-based epitaxial multilayers were grown in ultra-high-vacuum (UHV) condition on commercially available (111)-oriented oxide single crystals (MgO). The experiments were performed at IMDEA Nanoscience in a multipurpose UHV system combining MBE, DC-RF sputtering and XPS-UPS-LEED chambers connected (48). The MgO(111) crystals were ex-situ annealed in air at 1370 K for 2h in order to obtain flat surfaces with large terraces prior to their insertion in the UHV chamber. Epitaxial (111)-oriented Pt buffer with thicknesses ranging from 30 to 100 nm were deposited by DC sputtering in $8\times10^{-3}$ mbar Ar partial pressure at 670 K with a deposition rate of 0.3 Å/s. The quality of the fabricated Pt templates resembles the one of a single crystal, as demonstrated by LEED and XPS surface analyses.

The Pt/MgO(111) template was annealed at 1025 K in UHV ($1\times10^{-9}$ mbar) before the insertion of ethylene gas through a leak valve with a partial pressure of $2\times10^{-8}$ mbar for 30 min. After, the sample was cooled down to RT. Co was hence deposited on the top of gr/Pt/MgO(111) by e-beam evaporation at RT with a deposition rate of 0.04 Å/s. At each stage of the growth process, we have performed LEED and XPS measurements to check the electronic and chemical properties. The sample was gradually heated up to 550 K while acquiring XPS spectra with the aim to study in real time the intercalation of Co underneath the gr sheet, as well as the eventual intermixing between Co and Pt. The intercalation starts at 475 K and intermixing between Co and Pt occurs above 575 K. Once the intercalation was completed, the resulting structure was gr/Co($t$)/Pt//MgO(111). We have controlled the intercalation process as function of the amount ($t_{Co}$) of the evaporated Co, from 1 to 6 nm.

**XPS Measurements.** The XPS measurements were performed with unpolarized Al K$\alpha$ line (h$\nu$ = 1486.7 eV). The hemispherical energy analyzer (SPHERA-U7) pass energy was set to 20 eV for the XPS measurements to have a resolution of 0.6 eV. The angular acceptance for the used aperture size is defined solely by the magnification mode, i.e. $1750 \times 2750$ μm$^2$. The core level spectra were fitted to mixed Gaussian-Lorentzian components using CasaXPS software; all the binding energies are referred to the sample Fermi level. The best mixture of Gaussian-Lorentzian components is dependent on the instrument and resolution (pass energy), most of the measurements have been fitted with a line shape of GL(70) while for spectra with broad peak shapes and/or satellite structure line shapes of GL(30) are used for the individual components.

**High-Resolution TEM and EELS.** Electron microscopy observations were carried out in a JEOL ARM200cF microscope

equipped with a CEOS spherical aberration corrector and a Gatan Quantum EEL spectrometer at the Centro Nacional de Microscopía Electronica (CNME) at the University Complutense of Madrid. Specimens were prepared by conventional methods, including mechanical polishing and Ar ion milling. In-plane lattice distances where obtained from atomic resolution HAADF images by finding the atomic column positions using a center-of-mass approach, which allows a high degree of accuracy. For this aim, the substrate was used as internal calibration. Random noise was removed from EELS data using principal component analysis. EELS-derived compositional maps were produced both by subtracting the background using a power law fit followed by integration of the signal below the relevant edges except for the overlapping O $K$ and Pt $N_{2,3}$ edges. In this case a multiple linear least squares fit of the data was employed.

**Polar Kerr magnetometry and microscopy.** The RT vectorial-Kerr experiments were performed in polar configuration by using p-polarized light (with 632 nm wavelength) focused on the sample surface and analyzing the two orthogonal components of the reflected light. This provides the simultaneous determination of the hysteresis loops of the out-of-plane and in-plane magnetization components, i.e. $M_z$ and $M_y$, by sweeping the magnetic field along the sample out-of-plane ($\hat{z}$) direction. Details on the experimental vectorial Kerr setup can be found in ref. (49).

Kerr-microscopy experiments devoted to the study of domain wall (DW) dynamics were performed to determine both DMI strength and sign (DWs chirality) (28)(39). The experiment consists in measuring the DW velocity driven by an out-of-plane field, measured as a function of a constant in-plane field applied perpendicular to the domain wall direction. The $\mu_0 H_x$ field for which the velocity is minimum corresponds to the $\mu_0 H_{DMI}$ field that stabilizes chiral Néel walls. The strength of this field allows the determination of the DMI intensity. In order to carry out these experiments, the sample was first saturated with an out of plane magnetic pulse provided by an external electromagnet. An opposite magnetic-field pulse was then applied to nucleate some reverse domains. The pulses up to 600 mT were generated by a micro-coil with 200 μm diameter connected to a current source and a function generator capable of generating pulses down to 20 ns. The bubble domains expand from this first nucleation, by applying successive pulses. The velocity of the DW can be extracted by measuring such expansions, since the duration of the applied pulse is known.

In general, the DWs speed is limited by the breakdown that occurs at the Walker field, beyond which the DW starts precessing and the DW velocity drops. This happens in systems with negligible DMI where the DWs have achiral Bloch internal structure, as it was found in symmetric Pt/Co/Pt stacks (28)(39)(42). In this case, in fact, the two interfaces give rise to opposite DMI contributions that compensate each other. In asymmetric trilayers with strong DMI, e.g. when Co is embedded between two different non-magnetic heavy metals, the Walker field is pushed to higher fields so that large DW velocities can be obtained (19)(39).

**XAS-XMCD.** The XAS and magnetic circular dichroism experiments were carried out at the BOREAS beamline of the ALBA synchrotron using the fully circularly polarized X-ray beam produced by an apple-II type undulator (13). The base pressure during measurements was ~1×10$^{-10}$ mbar. The X-ray beam was focused to about 500 × 500 μm$^2$, and a gold mesh has been used for incident flux signal normalization. The XAS signal was measured with a Keythley 428 current amplifier via the sample-to-ground drain current (total electron yield TEY signal). The magnetic field was generated collinearly with the incoming X-ray direction by a superconducting vector-cryomagnet (Scientific Magnetics). To obtain the spin-averaged XAS and the XMCD, the absorption were measured as a function of the photon energy both for parallel and antiparallel orientation ($\mu^+(E)$ and $\mu^-(E)$) of the photon spin and the magnetization of the sample. We recall that such XMCD measurements at the Co and $L_{2,3}$ absorption edges provide direct element-specific information on the magnitude and sign of the projection of Co magnetizations along the beam and field direction. The projections of the orbital ($m_L$) and spin ($m_S$) magnetic moment along the incident light direction are extracted by using the XMCD sum rules (34)(38): $m_L = \frac{2}{3}\frac{n_h \mu_B}{P\cos\theta}\frac{\Delta A_3 + \Delta A_2}{A_3 + A_2}$ and $m_S = \frac{n_h \mu_B}{P\cos\theta}\frac{\Delta A_3 - 2\Delta A_2}{A_3 + A_2}$, with θ being the angle between the incident x-ray (θ=0º in perpendicular and θ=70º in grazing geometries) and sample magnetization direction (with circular polarization P=100%).

**Author Contributions**

P.P conceived the project. F.A., A.G., R.G. and P.P grew the samples. F.A., A.G., M. N., J.C. and P.P. acquired the data at the synchrotron with the help of P.G. and M.V. M.C. and M.V. acquired and analysed the STEM-EELs data. F.A., S.P. and J.V. acquired and analysed the Kerr microscopy and VSM-SQUID data. F.A., J.C., S.P. and P.P. treated and analysed the data. P.P. prepared the manuscript with the help of S.P., J.C., and R.M. All authors discussed and commented the manuscript.